
\magnification=\magstep 1
\tolerance 500
\rightline {TAUP 2134-94}
\vskip 3 true cm
\centerline {\bf The Unstable System}
\centerline{\bf in}
\centerline{\bf Relativistic Quantum Mechanics}
\vskip 1 true cm
\centerline {L.P. Horwitz \footnote{*}{Also at Department of
Physics, Bar Ilan University, Ramat Gan, Israel}}
\centerline{School of Physics}
\centerline{Raymond and Beverly Sackler Faculty
of Exact Sciences}
\centerline{Tel Aviv University, Ramat Aviv, Israel}
\vskip 2 true cm
\noindent {\it Abstract: \/}  A soluble model for the
 relativistically  covariant description of an unstable
system is given in terms of  relativistic quantum field
theory, with a structure similar to Van Hove's generalization
of the Lee model in the non-relativistic theory.  Since the
Fock space for this model can be decomposed to sectors,
 it can be embedded in a one-particle Hilbert
space in a spectral form  similar to the
 Friedrichs model in the non-relativistic theory.
Several types of spectral models result corresponding to
physically motivated assumptions made in the framework of
the field theory.  For example, the continuous spectrum
of the unperturbed problem may be $(-\infty, \infty)$ or
semibounded.  In the decay $V \rightarrow N + \theta$, the
$\theta$ may have negative energy.  In this case, the reaction
corresponds to the crossed channel process $V+ {\overline \theta}
\rightarrow N$.
\vfill
\eject
\noindent{\bf I. Introduction}
\par The Lee-Friedrichs model$^{1,2}$ has been very useful for the
study of the properties of unstable systems.  It is completely
soluble, and provides a framework for the analytic study of decay
and scattering systems$^3$.  The construction of generalized
states with exact exponential decay$^4$ in the rigged Hilbert
space$^5$ (Gel'fand triple) was originally motivated by the
properties of this model; such states have found application
in the theory of iterated maps$^6$ and play an important role
in the study of irreversible processes$^7$.  For both theoretical
development and  physical
applications, it is important to have a relativistic model for
unstable systems.  An unstable particle generally
decays into a final state of two or more particles in a process
for which the total energy is conserved, but the total
mass is not.   The equivalence between
mass and energy, which enters quantitatively in the kinematical
 description of such a process, is a fundamentally relativistic
relation.    We shall use a
formulation of relativistic quantum theory (and quantum field
theory, as will be discussed in Section 2) which is well-suited for
 this problem in that it treats
the energy and momentum of a particle as independent observables;
hence, the mass becomes a dynamical variable.  The transition from
the undecayed initial state to the decayed system in the final
state can then be studied as a continuous quantum mechanical
evolution.
\par This formulation$^8$, a generalization to the $N-$body system
of the relativistic quantum theory of Stueckelberg$^9$ developed
for a single particle, describes evolution with a Poincar\'e
invariant parameter $\tau$, sometimes called the ``historical
time.''  This invariant parameter
 corresponds to the universal time of (freely falling)
ideal clocks$^{10}$, and may be
identified with the universal time originally postulated
by Newton.  In this quantum theory, the four momenta $p^\mu$
are conjugate to the four coordinates of space and time $x^\mu
=(t, {\bf x})$. Classically, $t$, the time at which an event occurs in
the laboratory (measured on a standard ideal clock, i.e., running
with $\tau$), is a dynamical variable on the same footing as
${\bf x}$ (the place at which the event occurs in the laboratory).
The wave function, for example, for a one-particle
system, $\psi_\tau(x)$ (we shall use $x$ to represent $x^\mu$),
is the amplitude for the probability density $|\psi_\tau(x)|^2$ to
find an event at $(t,{\bf x})$ {\it at the historical time\/} $\tau$.
  Since this density is
 normalized with respect to integration over $t,{\bf x}$
(i.e., it belongs to the Hilbert space $L^2({\bf R}^4))$
 measurable unbounded excursions in $t$, as for ${\bf x}$ in the
non-relativistic quantum theory, are not admissible$^{11}$.
\par  Classically, the variables $t(\tau)$,
$ {\bf x}(\tau)$, correspond
to the value of $t, {\bf x}$ of a signal detected in the
laboratory (in some local inertial Lorentz frame) for
an event generated at time $\tau$ (in some, perhaps other, frame).
The result of the detection experiment is influenced by forces
 through the Hamilton equations ($i=1,2,\dots,N$
for an $N-$body system)
$$ {dx_i^\mu \over d\tau} = {\partial K \over \partial p_{i\mu}}
\qquad {dp_i^\mu \over d\tau} = - {\partial K \over
\partial x_{i\mu}}, \eqno(1.1)$$
where $K$ is a function on the $8N$ dimensional phase space,
as well as the state of motion of the frame by means of the
Lorentz transformation.
\par In the quantum theory, the Stueckelberg-Schr\"odinger
equation for an $N-$body system
$$ i {\partial \Psi_\tau \over \partial\tau}(x_1 \dots x_N) =
K(x_1 \dots x_N, p_1 \dots p_N) \Psi_\tau(x_1 \dots x_N)
\eqno(1.2)$$
describes the evolution of the wave function, and the dynamical
variables satisfy the Heisenberg equations
$$  {\dot x}^\mu \equiv {dx^\mu \over d\tau} = i [K,x^\mu]
\qquad {\dot p}^\mu \equiv {dp^\mu \over d\tau} = i [K,p^\mu],
\eqno(1.3)$$
with [$g^{\mu\nu} = (-,+,+,+)$;
 we shall use units for which $\hbar=
c=1$]
$$ [x^\mu, p^\nu ] = i g^{\mu\nu}. \eqno(1.4)$$
\par For a single free particle, the assumption
$$ K_0 = {p^\mu p_\mu \over 2M}, \eqno(1.5)$$
where $M$ is an intrinsic parameter belonging to the particle
which sets the scale of energy-momentum relative to its
space-time evolution in $\tau$, results in (both classically and
quantum mechanically)
 $$ {dx^\mu \over d\tau} = {p^\mu \over M}. \eqno(1.6)$$
For the classical case, one may eliminate $d\tau$ to obtain
$$ {d{\bf x} \over dt }= { {\bf p} \over E} , \eqno(1.7)$$
the Einstein relation for velocity. A localized ( in ${\bf x},t$)
free wave packet moves with $\tau$, according to the Ehrenfest
theorem, approximately along the classical world line.
 The generic form (one may also
include electromagnetic interactions in gauge invariant form)
$$ K = \sum_{i=1}^N {p_i^\mu p_{i\mu} \over 2M_i }
+ \ \ V(x_1 \dots x_N), \eqno(1.8)$$
where $V$ is a Poincar\'e invariant function of the $\{x_i^\mu\}$,
provides a family of consistent direct action potential models
to describe the evolution of a relativistic $N$-body system.
The Gibbs ensembles for the equilibrium relativistic statistical
mechanics of such systems, both for distinguishable and
indistinguishable particles has been worked out$^{12}$, as
well as the BBGKY and Boltzmann equations$^{13}$ for the
non-equilibrium evolution of the $N$-body system.  The equilibrium
limit of the Boltzmann distribution provides a relativistic
equilibrium {\it mass} distribution, for which the low energy
and non-relativistic limits have been investigated (with results,
in this limit, of course, in agreement with non-relativistic
statistical mechanics), and the relativistic Fermi and Bose
equilibrium distributions are currently being studied, for
example, for their consequences in astrophysics and condensed
matter theory$^{14, 15}$. Van Hove$^{16}$ has used the non-relativistic
Lee-Friedrichs model
to discuss some of the fundamental properties of non-equilibrium
statistical mechanics, and we shall, in a future publication,
apply the model that we have developed here to analogous
considerations.
\par As an illustration of the applicability of the model $(1.8)$,
 the two-body problem with ``central'' potential has been
studied in detail$^{17}$.  Since we shall make use of
two body kinematics in the study  of the decay problem, we
shall review here some of the details of the treatment of
this problem.  For the two-body case, we take
$$ K = {p_1^\mu {p_1}_\mu \over 2M_1} + {p_2^\mu {p_2}_\mu
 \over 2M_2} + V(\rho),      \eqno(1.9)$$
where $\rho = \sqrt{x^\mu x_\mu}$, and the relative space-time
coordinates are given obey
$$ x^\mu = x_1^\mu - x_2^\mu . \eqno(1.10)$$
In the non-relativistic limit, $t_1 \rightarrow t_2$, so that
$\rho \rightarrow r = \vert {\bf x}_1 - {\bf x}_2 \vert$.
The total momentum of the system is
$$ P^\mu = p_1^\mu + p_2^\mu ,  \eqno(1.11)$$
which commutes with the relative coordinate $x^\mu$.  The
operator $K$, as for the non-relativistic analog, can therefore
be written as
$$ K = {P^\mu P_\mu \over 2M} + K_{rel}, \eqno(1.12)$$
where
$$ \eqalign{K_{rel} &= {p^\mu p_\mu \over 2m}
+ V(\rho) , \cr p^\mu &= { M_2p_1^\mu - M_1p_2^\mu \over
{M_1 + M_2}} , \cr}
\eqno(1.13)$$
corresponds to the evolution operator for the relative motion.
Clearly, $K$ has only continuous spectrum, since the
center of momentum motion is that of a free particle.  Defining
$$ X^\mu = {{M_1 x_1^\mu + M_2 x_2^\mu} \over M}, \eqno(1.14)$$
where
$$ M= M_1 + M_2 , \eqno(1.15)$$
we see that by $(1.1)$,
$$ {dX^\mu \over d\tau} = {P^\mu \over M}.  \eqno(1.16)$$
However, $K_{rel}$, of $(1.13)$, can have discrete or continuous
spectrum in spite of the fact that $p^\mu p_\mu$ is a hyperbolic
operator.  One studies this problem by separating the center
of momentum variable from $K$ (i.e., taking a direct integral
representation of the solution over $P^\mu$, the absolutely
conserved total energy-momentum).  It is a remarkable fact that
if we take the support of the eigenfunction $\psi_k(x)$, of the
equation
$$ K_{rel} \psi_k(x) = k \psi_k(x)  \eqno(1.17)$$
to lie in the sub-measure space defined by
$$ x_1^2 + x_2^2 - t^2 \geq 0, \eqno(1.18)$$
called the RMS (restricted Minkowski space), the spectrum of
$K_{rel}$  then coincides with the spectrum of the
non-relativistic Schr\"odinger equation for all central potentials
$V(r) = V(\rho)|_{\rho = r}$.  The corresponding energies are
then determined by $(1.12)$, i.e., in the center of momentum
frame, for which ${\bf P} = 0$,
$$E= \sqrt{2M(k - K)}.\eqno(1.19)$$
With the assumption that the particles have masses close to $M_1,
M_2$ when they are asymptotically spacelike separated (for which
$V \approx 0$),
 and that the
corresponding evaluation of $K$  is
constant over the entire spectrum of $K_{rel}$ [we argue that
$t$-dependent perturbations of $V$ at long intervals can shift the
spectrum of $K_{rel}$, but leave the absolutely conserved $K$
constant], $(1.19)$ becomes
$$ E \cong  \sqrt{2Mk + M^2} . \eqno(1.20)$$
For the case of discrete spectrum, where $k=k_n$, $n$ integer,
when the excitations are small in magnitude compared to $M$,
$$ E_n \cong M + k_n + MO(({k_n \over M})^2 ). \eqno(1.21)$$
The same decompostion can be carried out for continuum values of
$k$ (scattering states) small compared to $M$.
  \par The RMS has a
natural coordinatization in terms of the compact parameters
$\theta, \phi$ and the non-compact parameter $\beta$ (along
with $\rho$) as
$$ \eqalign{ x_1 &= \rho \cos \phi \sin \theta \cosh \beta \cr
x_2 &= \rho \sin\phi \sin\theta \cosh\beta \cr
x^0= t &= \rho \sin\theta \sinh\beta \cr
x_3 &= \rho \cos\theta ; \cr} \eqno(1.22)$$
clearly, $ x_1^2 + x_2^2 - t^2 = \rho^2 \sin^2 \theta \geq 0$,
and $x_1^2 + x_2^2 +x_3^2 - t^2 = \rho^2 \geq 0$ as well.
This submeasure space is $O(2,1)$ invariant, but not
$O(3,1)$ invariant.  The representations of $O(3,1)$ are
induced from the irreducible representations of $O(2,1)$
which are provided by the eigenfunctions of $(1.17)^{17}$.
\par A model for relativistic hydrogen is obtained by taking
$V=-e^2/\rho$.  The exact eigenfunctions are therefore known
for this problem; it has recently been shown that the selection
rules for radiation coincide with those predicted by the
non-relativistic theory$^{18}$, and a relativistic form
of the Zeeman effect has been worked out$^{19}$.  Many
interesting group theoretical problems have been posed by
this structure.  The dynamical group for the relativistic
Kepler problem has been found$^{20}$, and a general investigation
of such non-compact dynamical groups is being carried out$^{21}$.
\par The theory of unstable systems that forms the basis of our
discussion in this work was first given in a systematic way
by Wigner and Weisskopf$^{22}$, in their work on the spectral
line width for atomic radiation.  In this (non-relativistic)
theory, the unstable
 system is represented  by a wave function $\psi$ which is usually
taken to be an eigenfunction of an
 unperturbed Hamiltonian $H_0$; the
action of the full perturbed Hamiltonian $H$
 then causes a non-trivial
evolution of the wave function $\psi_t = e^{-iHt} \psi$,
 and the amplitude
$$ A(t) = (\psi, e^{-iHt} \psi)    \eqno(1.23)$$
is understood to be the amplitude for survival of the initial state,
i.e.,
$$ p(t) = \vert A(t)\vert^2 \eqno(1.24)$$
is the probability to find the initial state ``undecayed''
after a time $t$.  After a time $t$, not too short and not too
 long, the evolution $(1.23)$ is, for a wide class of models,
well-approximated by the exponential of a complex number, i.e.,
of the form $\exp\{-i(E_0 - i{\gamma \over 2})t\}$, where $\gamma$
is the exponential decay rate for $p(t)$.  If $H$ is defined
on $\psi$, it is well known that the survival probability
$p(t)$ has zero derivative at $t=0$, and, moreover, it goes
asymptotically, for large $t$, as an inverse power$^{23}$ of $t$.
\par It should be remarked in this connection that a fundamental
assumption underlying the Wigner-Weisskopf theory is that,
under continuous evolution,
$$ e^{-iHt} \psi = A(t) \psi + \psi^\perp,$$
where $\psi^\perp$ is the part of $e^{-iHt}\psi$ orthogonal
to $\psi$.  The linear superposition $A(t)\psi + \psi^\perp$ is,
in fact, not seen in the laboratory.  One generally sees only the
state $\psi$ or the decay products.  It appears that the time of
decay is a superselection rule, and that this linear superposition
does not correspond to the physical situation.  This problem
has been discussed by Horwitz and Piron$^{24}$, where an
alternative formulation of the problem for the non-relativistic
case has been given.  We shall not discuss the relativistic
generalization of that treatment here, but confine ourselves
to the usual Wigner-Weisskopf form which, although perhaps not the
most general theoretical framework, has led to a good quantitative
approximation for the non-relativistic theory.
\par One may understand the time evolution predicted by the Wigner-
Weisskopf theory most easily by examining the inverse Laplace
transform$^{25}$ (one takes $A(t) = 0 $ for $t<0$)
$$ A(t) = {1 \over 2\pi i} \int_ C \ dz
 (\psi, {1 \over {z-H}}
\psi) e^{-izt}           \eqno(1.25)   $$
where $C$ is a contour that depends on the nature of the spectrum
of $H$.
\par If $H$ has continuous spectrum, for example, on $(0,\infty)$,
then the function
$$ R(z) = (\psi, {1 \over {z-H}}\psi)  \eqno(1.26)$$
is analytic in the upper half plane along with its extension
in the first Riemann sheet to the lower half plane, excluding
the cut along the positive real axis.  The contour $C$ then
corresponds to a line running above the positive real axis
from $\infty$ to $0$, then going around the branch point and
continuing from $0$ to $\infty$ below the real axis.  The
lower part of this contour can be deformed to a vertical line
from $0$ to $-i\infty$ for $t>0$, providing only
 a small contribution for large $t$.  The part of the contour
lying above the real axis may then be deformed through the cut
 to the second Riemann sheet, lowering it eventually to a line
in the second sheet from $-i\infty$ to $0$, going around the
branch point to connect with the part of
contour that remains in the
first sheet.  In the course of this deformation, the contour passes
the pole which has been shifted (for small coupling) from the
discrete (bound) initial state embedded in the continuum on the
real axis into the second sheet in the lower half plane, and this
pole and its residue, proportional to $e^{-i\zeta t}$, where
$\zeta$ is the position of the pole, corresponds to the
 exponential decay law.  For large $t$, this contribution becomes
small compared to that provided by the neighborhood of the branch
point, which yields an inverse power of $t$.  For very short times,
the contribution of the deformed integral cannot be ignored, and,
in fact, the integral is best approximated along the real
 axis$^{26}$.  Since $H$ is self-adjoint, clearly $dp(t)/dt$
vanishes at $t=0$, and it is easily seen that the power series
in $t$ starts as $1-O(t^2)$.
\par Friedrichs$^2$ proposed the following soluble model, which
provides a closed analytic form for what is called the reduced
resolvent $R(z)$, for which the analysis described above may
be carried out in a very simply.  Let
$$ H= H_0 + V,     \eqno(1.27)$$
where $H_0$ has an absolutely continuous spectrum in $(0, \infty)$
and a single bound (discrete) state $\psi$ for which
$$ H_0 \psi = E_0 \psi.  \eqno(1.28)$$
Furthermore, suppose that on the continuous spectrum of $H_0$
characterized by
$$ \langle E \vert H_0 \vert f) = E \langle E \vert f),
 \eqno(1.29)$$
for any $f$ in the domain of $H_0$, and that
$$ \langle E \vert V \vert E' \rangle = 0 \eqno(1.30)$$
for all $E, E'$.  The diagonal part $(\psi \vert V \vert \psi)$
of $V$, leading to a simple shift in $E_0$, is usually taken to
be zero.  Then, it follows from the identity (second resolvent
equation)
$$ {1 \over {z-H}} = {1 \over {z-H_0}} + {1 \over {z-H_0}}V
{1 \over {z-H}}      \eqno(1.31)$$
that
$$R(z) = {1 \over {z-E_0}} + {1 \over {z-E_0}} \int dE \ (\psi
\vert V \vert E \rangle \langle E \vert {1 \over {z-H}} \vert
\psi)   \eqno(1.32)$$
and
$$\langle E \vert {1 \over{z-H}}\vert \psi) = {1 \over {z-E}}
\langle E \vert V \psi) R(z).   \eqno(1.33)$$
\par The simplifying truncation in $(1.33)$ follows from the
essential feature $(1.30)$ of the model, implying physically
that there are no ``final state'' interactions.  This
is often, in fact, a good physical approximation in decaying
systems.  Substituting $(1.33)$ into $(1.32)$, one finds
$$ h(z) R(z) = 1, \eqno(1.34)$$
where
$$ h(z) = z - E_0 -\int dE\ {\vert \langle E \vert V \vert \psi)
\vert^2 \over {z - E} } .  \eqno(1.35)$$
\par The singularities of $R(z)$ arise from the vanishing of
the function $h(z)$.  Since
$$ {\rm Im}H(z) = {\rm Im}z \biggl[1 + \int dE\ {\vert \langle E
\vert V \vert \psi ) \vert^2 \over \vert z- E\vert^2 } \biggr],
\eqno(1.36) $$
there can be no zero in the first sheet of the function for
$ {\rm Im}z \neq 0$. For sufficiently small $V$, it is easy to
show that there is no zero on the negative real axis.  Passing
to the second sheet continuously through the cut, one finds$^{25}$
$$ h^{II} (z) = h(z) + 2 \pi i W(z), \eqno(1.37)$$
where $W(z)$ is the analytic extension of
$$ W(E) = \vert \langle E \vert V \vert \psi) \vert^2 $$
to a region of the lower half plane sufficiently large [that is,
we assume an analytic function in this domain whose boundary
is $W(E)$] to include the possible zero of $h^{II}(z)$.  If this
zero occurs for $z=\zeta$ with small imaginary part, then
$$ Im \zeta \biggl[ 1 + \int dE\ {\vert \langle E \vert V \vert
\psi) \vert^2 \over \vert z - E \vert^2 } \biggr] \cong
- 2 \pi \vert \langle E_0 \vert V \vert
 \psi)\vert^2, \eqno(1.38)$$
so that, recalling the representation
$$ {\lim_{\epsilon \rightarrow 0}} {1 \over \pi} {\epsilon
\over {\epsilon^2 + x^2}} = \delta(x) ,$$
one sees that the pole occurs at
$$ {\rm Im} \zeta \cong - \pi \vert \langle E_0 \vert V \vert
\psi)\vert^2.  \eqno(1.39)$$
\par It is clear that for the description of, for example, particle
decay, involving a real change in the the total particle mass of
the system, a relativistic theory is required. Since, as
pointed out by Henley and Thirring$^{27}$, in
the standard on-shell relativistic
quantum field theory, both the annihilation
operator for a particle and the creation operator for an antiparticle
occur linearly combined in the local field operator, a relativistic
analog of the model described above (in the framework of quantum
field theory, in the form given by Lee$^1$) would contain,  along
with the process $ n \leftrightarrow p + \pi^- $,
automatically and simultaneously, the process
$ p \leftrightarrow n + \pi^+ $.  Zumino$^{28}$ has
remarked that the expression $E= \sqrt{ {\bf p}^2 + m^2}$ is not
analytic, and creates singularities that make the analytic
continuations we have discussed above difficult;
 he suggested,
toward the end of his article, the use of the quadratic form
$p^\mu p_\mu$ instead, a form which we will, in fact, utilize in the
formulation we give in the next sections. In the field theoretic
form, the ``coupling'' which plays the role of $\langle E\vert V
\vert \psi)$, is a function of the momentum of one of the particles
emitted in the decay; its Fourier transform to configuration space
results in a non-local function in space which would make
Lorentz invariance difficult to achieve.  The original form
of the model proposed by Lee$^1$ , which was constructed
to achieve the simplest
form, did not allow for recoil,
and hence its relativistic generalization
would destroy Poincar\'e invariance; however, in the form given
by Van Hove$^{29}$, recoil is taken into account.  It is for this
 form of the field theory which we shall provide a
relativistic generalization.
\par It is the purpose of this paper to show how such a
relativistically covariant version of this model can be
constructed.  In Section 2, we shall formulate the decay problem
in terms of the relativistic quantum field theory resulting
from the second quantization of
 the covariant quantum theory we have
discussed above.  It has recently been shown that such a field
theory results in a consistent quantum electrodynamics$^{30}$
involving a five dimensional generalization of the Maxwell field.
This field (including a Lorentz scalar
 field as the fifth component)
 reduces to the usual
Maxwell field when a characteristic correlation length reaches
a critical value.  The theory has been shown$^{31}$ to provide
insight into regularization (such as that of Pauli-Villars$^{32,33}$)
and renormalization methods used in standard quantum electrodynamics
(e.g., refs. 34).  We shall not describe here this general theory,
but use the relativistic matter fields of this theory in much the
same way that Lee$^1$ and, in particular, Van Hove$^{29}$
used the non-relativistic quantized fields to describe the
soluble Lee model.  The information we obtain from this field
theory, in a model which has two conserved quantum numbers
defining particle number {\it sectors},
 is then used, in Section 4, to construct a corresponding
model in a single Hilbert space (which is not a tensor product
Fock space representing a quantum field theory) according to
the method of Friedrichs$^2$.  We thus achieve
a relativistically covariant Lee-Friedrichs soluble model
for an unstable system.
\par It is a pleasure to dedicate this paper to Professor
Jean-Pierre Vigier on the $45^{th}$ anniversary of his
joining Louis de Broglie at the Institut Henri Poincar\'e.
The theory of E.C.G. Stueckelberg provides a framework in
which the notion of de Broglie$^{35}$, that the mass of a
particle is associated with a quantum frequency, is realized
through the identification of the energy in a relativistic
theory as an observable kinematically independent
 of the momentum.  The work of Vigier and collaborators (ref. 8),
contributing to the development of this theory, we feel, makes
the present work appropriate for such an occasion.
\bigskip
\noindent
{\bf 2. Field Theory Model}
\smallskip
\par In order to determine the structure and physically motivated
basic assumptions to be imposed on a Friedrichs type model for the
relativistic description of an unstable system in a ``one-
particle'' Hilbert space, we first study in some detail a model
in relativistic quantum field theory.  We shall assume boson
matter fields in what follows.  The properties of the quantum
fields may be obtained by a variety of methods; one of these is the
construction of a Lagrangian density for which the variation of
the associated action yields a field equation which coincides with
the Stueckelberg-Schr\"odinger equation.  A Lagrangian density
which yields this result is, for example$^{30}$,
$$ \eqalign{{\cal L} &= {1 \over 2} \{i\psi_\tau^\dagger \partial_\tau
\psi_\tau - \partial_\mu\psi_\tau^\dagger \partial^\mu \psi_\tau
\ \ \ + \ \ \ \rm c.c.\} \cr
&- \int dx\ dx'\ \psi_\tau^\dagger (x) \psi_\tau^\dagger (x') V(x-x')
\psi_\tau(x')\psi_\tau(x)  \}, \cr} \eqno(2.1)$$
and the action is defined as
$$ S = \int d\tau \int d^4x\ {\cal L}. \eqno(2.2)$$
\par It then follows that the field
 conjugate to $\psi$ is $\Pi \equiv
i\psi^\dagger$, so that the equal-$\tau$ commutation relations
are
$$ [ \psi_\tau(x), \psi_\tau^\dagger(x') ] = \delta^4(x-x').
\eqno(2.3)$$
In momentum space, for which
$$ \psi_\tau(p) = { 1 \over (2 \pi)^2} \int d^4x\ e^{-ip^\mu x_\mu}
\psi_\tau(x),  \eqno(2.4)$$
this relation becomes
$$ [ \psi_\tau(p), \Pi_\tau(p') ] = i\delta^4(p-p').
\eqno(2.5)$$
\par To see how this relation goes over to the on-mass-shell
field theory, we follow a non-rigorous but instructive
procedure.  Multiply both sides of $(2.5)$ by $\bigtriangleup E$,
and notice that for $E= \sqrt{{\bf p}^2 + m^2}$, where $m^2$ is
a dynamical variable, $\bigtriangleup E
 = \bigtriangleup m^2/2E$, we obtain
$$[ {\tilde \psi}_\tau ({\bf p}), {\tilde \Pi}(p') ] =
2iE\delta^3({\bf p} - {\bf p}'), \eqno(2.6)$$
  where ${\tilde \psi}_\tau({\bf p}) = \sqrt{\bigtriangleup m^2}
\psi_\tau(p)$, and ${\tilde \Pi}_\tau ({\bf p}) = \sqrt
{\bigtriangleup m^2} \Pi_\tau(p)$.  We may then take the limit
$\bigtriangleup m^2 \rightarrow 0, m^2 \rightarrow m_0^2$, a
fixed constant (in this limit, $t \rightarrow \tau$ as well$^8$).
The operator-valued distribution $\psi_\tau(p)$ therefore peaks
 in support
(in all matrix elements) sharply at $m^2 \sim m_0^2$ in this limit.
Eq. $(2.6)$ is the conventional equal time commutation relation.
\par The example we have studied above assumes positive energy.
Both fields $\psi, \Pi$ may have negative energy as well, but the
right hand side vanishes if one has positive,  and the other
negative ($\bigtriangleup E \delta(E-E')$ would be zero in this
case).  We remark, furthermore, that according to $(2.5)$, the field
$\psi(x)$ contains {\it only} the annihilation operator
 $\psi_\tau(p)$, for both positive and negative energy, but
not a creation operator for the antiparticle (as for the second
quantization of the Schr\"odinger wave function).  The CPT
conjugate of a negative energy function corresponds to the
representation of the antiparticle observed in the laboratory.  This
is consistent with Stueckelberg's (and Feynman's)$^9$ original
interpretation, associating the antiparticle with the
particle going ``backwards'' (with $\tau$) in time.  The
quantum electrodynamics (a $U(1)$ gauge
 field in interaction with the
matter field) associated with this theory is obtained by requiring
gauge invariance of the Stueckelberg equation$^{30}$.
In addition to the derivative $-i\partial^\mu$, which
becomes the gauge covariant
 derivative $-i\partial^\mu -a^\mu(x,\tau)$, the
  derivative $i\partial_\tau$ also requires compensation,
i.e., $i\partial_\tau \rightarrow i\partial_\tau
+ a_5(x,\tau)$, introducing a fifth, Lorentz scalar, field.
\par  It has been shown that the Maxwell potential (satisfying the
usual Maxwell equations with a four dimensional conserved current
as its source) is obtained from the field $a^\mu(x,\tau)$
by integration over $\tau^{30}$, i.e., the Maxwell field is
the zero mode of the ``pre-Maxwell'' field (the $a^5$ field
decouples in the $U(1)$ theory).  In the limit in which the
Fourier transform $a^\mu(x,s)$
 on $\tau$ of the pre-Maxwell fields
have support only in a small interval $\bigtriangleup s$ near zero
(a long correlation length limit) one can show that the theory
goes over to the usual form of quantum electrodynamics$^{30}$.
The theory has been consistently quantized, using canonical
methods$^{36}$, in full generality. In the quantized form, the
theory goes over to the standard quantum electrodynamics,
carrying
Pauli-Villars regularization$^{33,34}$, in the long correlation
length limit,
as we have remarked above.
We shall not discuss this structure further here, but concentrate
on a model field theory with a generalized invariant Hamiltonian
of the form
$$ K = K_0 + K_I  \eqno(2.7)$$
where (we write $p^2 \equiv p^\mu p_\mu$
 , $k^2 \equiv k^\mu k_\mu$
henceforth)
$$ \eqalign{K_0 &= \int d^4p {p^2 \over 2M_V} b^\dagger(p) b(p)
\cr &+ \int d^4p {p^2 \over 2M_N} a^\dagger_N(p) a_N(p)
\cr &+ \int d^4k {k^2 \over 2M_\theta}
a^\dagger(k)_\theta a_\theta(k), \cr} \eqno(2.8)$$
and
$$\eqalign{K_I&= \int  d^4k\ d^4p\ \{ f(k) b^\dagger(p) a_N(p-k)
a_\theta(k) \cr &+ f(k)^*a^\dagger_N(p-k) a^\dagger_\theta(k)
b(p) \}, \cr}\eqno(2.9)$$
describing the process $ V \leftrightarrow N + \theta $.  Here,
$b(p)$ is the annihilation operator for the $V$ particle, and
$a_N(p),\  a_\theta(k)$, for the $N$ and $\theta$ respectively.
This model is a relativistic generalization of Van Hove's
form$^{29}$ of the Lee model for the non-relativistic
theory (note that expressing the ``kinetic energy'' in the
on-mass-shell form $\sqrt{{\bf p}^2 + m_0^2}$ would make it
impossible to separate the center of momentum motion in the
two-body channel).  The operators
$$ \eqalign{Q_1 &= \int d^4p\ \bigl[b^\dagger(p)b(p)
 + a_N^\dagger(p) a_N(p) \bigr]\cr
      Q_2 &= \int d^4p\ \bigl[ a_N^\dagger(p) a_N(p) -
a_\theta^\dagger(p) a_\theta(p) \bigr] \cr}\eqno(2.10)$$
are conserved (as one easily sees using the commutation relations
$(2.5)$), and enable us to decompose the Fock space to sectors.
We shall study this problem in the lowest sector $Q_1=1,\  Q_2= 0$
for which there is just one $V$ {\it or} one $N$ and one $\theta$.
\par We now wish to calculate
$$ \Psi_\tau = e^{-iK\tau} \Psi_0, \eqno(2.11)$$
where we take the normalized wave packet
$$ \Psi_0 = \int g(p) b^\dagger(p) d^4p \vert 0 \rangle
\eqno(2.12)$$
as the one-particle initial state (the $V$-particle).  We can
now compute the relativistic analog of the survival amplitude
$$ A_s(\tau) = \bigl(\Psi_0, \Psi_\tau\bigr) =
\bigl(\Psi_0, e^{-iK\tau} \Psi_0 \bigr) \eqno(2.13)$$
as a function of the invariant time $\tau$.
The initial state may be chosen
 so that $\tau$ is the mean proper time for the
undecayed system (i.e., for which $\langle (dx^\mu/d\tau)\rangle
\langle (dx_\mu/ d\tau)\rangle = -1$).
\par Since $A(\tau)$ is Lorentz invariant,
we may transform from
an arbitrary frame to the (average) rest frame of the initial
state under the evolution induced by $K$.  Its $\tau$-dependence
therefore clearly can arise only from the evolution of $\Psi_0$
to a different channel (assumed here to be more rapid than the
significant spreading of the wave packet).
\par In the sector $Q_1=1$ and $Q_2 = 0$, conserved by the
evolution, a general representation for the Fock space state
$\Psi_\tau$ is
$$ \Psi_\tau = \int d^4p A(p,\tau) b^\dagger(p) \vert 0 \rangle
+ \int d^4p d^4k \ B(p,k,\tau) a_N^\dagger(p) a_\theta^\dagger(k)
\vert 0 \rangle , \eqno(2.14)$$
where we choose for the initial condition
$$ \eqalign{A(p,0) &= g(p) \cr B(p,k,0) &= 0, \cr} \eqno(2.15)$$
so that the initial state is that given by $(2.12)$.  The
evolution of the system is given by
$$ i {\partial \Psi_\tau \over \partial \tau} = K \Psi_\tau,
\eqno(2.16)$$
the quantum field theory form of Eq. $(1.2)$.  Applying the
definitions $(2.8), (2.9)$ for $K$ and the commutation relations
for the annihilation-creation operators, one obtains the
differential equations
$$ i {\partial A \over \partial \tau}(p,\tau) = {p^2 \over 2M_V}
A(p,\tau) + \int d^4k f(k) B(p-k,k,\tau) \eqno(2.17)$$
and
$$ i{\partial B \over \partial \tau} (p,k,\tau) = \biggl(
{p^2 \over 2M_N} + {k^2 \over 2M_\theta} \biggr) B(p,k,\tau)
+ f^*(k)A(p+k, \tau). \eqno(2.18)$$
Defining the Laplace transforms
$$ \eqalign{{\tilde A}(p,z) &=
 \int_0^\infty e^{iz\tau} A(p,\tau) d\tau \cr
{\tilde B}(p,k,z) &= \int_0^\infty e^{iz\tau} B(p,k,\tau), \cr}
\eqno(2.19)$$
we obtain from $(2.17),\ (2.18)$ the relations
$$ \biggl( z - {p^2 \over 2M_V}\biggr) {\tilde A}(p,z)
= ig(p) + \int d^4k f(k) {\tilde B}(p-k,k,z), \eqno(2.20)$$
where the term $ig(p)$ arises from the boundary point at $\tau = 0$
of the integration by parts (the contribution at $\tau \rightarrow
\infty $ vanishes for $z$ in the upper half plane), and
$$ z{\tilde B}(p,k,z)= \biggl( {p^2 \over 2M_N} + {k^2 \over
2M_\theta} \biggr) {\tilde B}(p,k,z) + f^*(k) {\tilde A}(p+k,z).
\eqno(2.21)$$
Substituting $(2.21)$ into $(2.20)$, we obtain the covariant analog
of Eq. $(1.34)$,
$$  h(p,z) {\tilde A}(p,z) = ig(p), \eqno(2.22)$$
where
$$ h(p,z) = z - {p^2 \over 2M_V} - \int d^4k {\vert f(k)\vert^2
\over z - {(p-k)^2 \over 2M_N} - {k^2 \over 2M_\theta}}.
\eqno(2.23)$$
\par In this form, one recognizes the essential content of the
relativistic Lee model.  The term $p^2 / 2M_V$ corresponds to
the mass (squared) of the unstable system, i.e., the discrete
(for fixed $p^\mu$) eigenvalue of the unperturbed initial state,
and the cut, the singularity corresponding to the continuous
spectrum of the decay channel.
\par The Laplace transform of the survival amplitude $(2.13)$,
$$ {\tilde A}_s(z) = \int_0^\infty e^{iz\tau} A_s(\tau) d\tau
= i\biggl( \Psi_0, {1 \over z-K} \Psi_0 \biggr), \eqno(2.24)$$
analytic for $z$ in the upper half-plane, is then given, according
to $(2.12)$ and $(2.14)$, by
$$ {\tilde A}_s(z) = \int d^4p g(p)^*
 {\tilde A}(p,z), \eqno(2.25)$$
i.e.(with $(2.22)$),
$${\tilde A}_s(z) = i \int d^4p {\vert g(p)\vert^2 \over h(p,z)}.
\eqno(2.26) $$
The inverse transform is
$$ A_s(\tau) = {1 \over 2\pi i} \int_C  e^{-iz\tau}
{\tilde A}_s (z) dz, \eqno(2.27)$$
where $C$ is a contour taken from $+\infty +i\epsilon$ to
$-\infty + i\epsilon$, i.e., just above the real axis.  In the case
of semi-bounded spectrum for the continuum channel, for $\tau
>0$, the left hand part of the contour can be deformed
 to the lower half-plane, where this part of the contour
(except for the neighborhood of the branch point) provides
a small background contribution for large $\tau$.  In the
same way as discussed in Section 1, the right hand half line
can be rotated into the second sheet, and brought down to
the line $(0,\infty)$, picking up the contribution of the
pole of $h(p,z)$ on the way.  If the spectrum is unbounded,
analytic continuation can be made by bringing the whole line
of integration down into the lower half plane.
\par Note that the contribution of the pole corresponding to the
zero of the analytic continuation of $h(p,z)$ is spread out
by the integration $(2.26)$, but this spreading can be very
small (it occurs primarily in the real part of the pole, since
the imaginary part, for weak coupling,
 is almost independent of $p$).  In the next
section, we discuss the structure of the spectrum of the
 continuous channel.
\bigskip
{\bf 3. The Spectrum}
\smallskip
\par Clearly, the unperturbed spectrum (for $f(k)=0$) for the
relativistic Lee model, according to $(2.8)$, is purely
continuous.  We see, however, from the structure of $(2.23)$
that for {\it each value} of $p^2$, there is an effective discrete
eigenstate with eigenvalue $p^2/2M_V$, considering
the Hilbert space as a direct sum space over the absolutely
conserved total energy-momentum, which coincides with $p^\mu$,
of the system. The remaining terms in $(2.8)$ can be written
as
$$ \int d^4k\ d^4p \biggl\{{(p-k)^2 \over 2M_N} +
{k^2 \over 2M_\theta }\biggr\}a_N^\dagger(p-k)a_\theta^\dagger(k)
a_\theta(k) a_N(p-k),  \eqno(3.1)$$
since we can replace $p-k$ by $p$; the integration of the first
term over $k$ then provides a factor which is the number operator
of the $\theta$ field, and the second term has a factor, after
 integration over $p$, which is the number operator for the
$N$ field.  In the sector $Q_1 = 1$ and $Q_2 = 0$, these factors
are unity in the continuum channel and zero in the initial state
channel, and hence do not change the valuation of this
 contribution to the unperturbed spectrum.  Clearly, $(3.1)$
corresponds to the continuous spectrum associated with the cut
in $(2.23)$, for each fixed $p^\mu$.  We therefore study the
range of the function
$$   w(k) =  {(p-k)^2 \over 2M_N} + {k^2 \over 2M_\theta}\ \ .
\eqno(3.2).$$
\par The range of the contribution of the continuous
 spectrum to $h(p,z)$
in $(2.23)$ is determined by the support properties of the
coupling function $f(k)$; it may contain, in an invariant way,
restriction to $k^0$ positive or negative, or bounds on
$m_\theta^2 \equiv -k^\mu k_\mu$, the mass-squared of the $\theta$
particle.
\par To study these cases, we write  $(3.2)$ as
$$ w(k) = {p^2 \over 2M_N} -{m_\theta^2 \over 2}\bigl({1 \over
M_\theta}+ {1 \over M_N} \bigr) + {1 \over M_N }( \pm p^0
\sqrt{{\bf k}^2 + m_\theta^2} -
 {\bf p} \cdot {\bf k}), \eqno(3.3)$$
where the signs $\pm$ in $(3.3)$ correspond to the choice of
sign of the energy variable of the $\theta$ particle, i.e.,
$$ k^0 = \pm \sqrt{{\bf k}^2 + m_\theta^2}. \eqno(3.4) $$
The choice of sign determines the direction of propagation of the
wave packet of the $\theta$-particle in time.  As pointed out by
Stueckelberg (and Feynman)$^9$, the negative energy particle
has a correspondence with the antiparticle;
the $CPT$ conjugate of the wave
packet is that of a particle with {\it positive} energy
and the opposite sign of the momentum and the charge, i.e., the
antiparticle.  Applying this definition to a particle with
spin$^{37}$, one finds that it is completely consistent with Dirac's
interpretation (the representations of $C, P$ and $T$ on the
corresponding Dirac spinors are the same as in the usual on-shell
theory)$^{38}$.  We first discuss the case $k^0 >0$.
\par We shall take $p^0 > 0$ in the following;
 the $\theta$ and $N$  particles can exchange roles under a slight
reformulation, and there is therefore a symmetry in the analysis
for $p^0 <0$.  Then since $p^0 > \vert {\bf p}\vert$, if
the $V$ particle has positive mass-squared (this condition
follows from the choice of states of the quantum field), it
follows from $(3.3)$ that
$$ w(k) \rightarrow +\infty \ \ \  {\rm for}\ \ \  \vert{\bf k}\vert
\rightarrow \infty.   \eqno(3.5)$$
The lower bound of the spectrum is determined, for each $p^\mu,
m_\theta$, as a function of $\vert {\bf k} \vert$, by
$$ {dw(k) \over d\vert {\bf k} \vert} = 0.$$
Let us define the ${\bf k}$-dependent part of $(3.3)$ as
$$ \eta_+ ({\bf k})= p^0 \sqrt{ {\bf k}^2 + m_\theta^2}
- \vert {\bf p}\vert \vert {\bf k} \vert \cos\,\phi, \eqno(3.6)$$
where $\phi$ is the angle between ${\bf p}$ and ${\bf k}$.  Then,
this condition implies
$$ \eta'_+({\bf k}) = { p^0 \vert {\bf k} \vert \over
\sqrt {{\bf k}^2 + m_\theta^2}} - \vert {\bf p} \vert \cos\,\phi .
\eqno(3.7)$$
The minimum value is therefore (the second derivative is positive)
achieved for $\cos\,\phi >0$, at
$$ {p^0}^2 {\bf k}^2 = {\bf p}^2 ({\bf k}^2 + m_\theta^2)\cos^2\,\phi,
\eqno(3.8)$$
yielding
$$ \vert {\bf k}\vert = {\vert {\bf p} \vert \cos\,\phi
\over \sqrt {{p^0}^2 - {\bf p}^2\cos^2\,\phi}}
 m_\theta , \eqno(3.9)$$
so that
$$ \sqrt{{\bf k}^2 + m_\theta^2} = {m_\theta p^0 \over
\sqrt{{p^0}^2 - {\bf p}^2 cos^2\,\phi}}.$$
It then follows that
$$ \bigl( \eta_+({\bf k}) \bigr)_{min} = m_\theta
\sqrt{{p^0}^2 -{\bf p}^2
\cos^2\,\phi} ,   \eqno(3.10)$$
which is further minimized by taking $\phi = 0$:
$$ \bigl(\eta_+({\bf k}) \bigr)_{min} = m_\theta \sqrt{-p^2} .
\eqno(3.11) $$
We note that for $\cos\,\phi < 0$, the minimum value is at
$p^0 m_\theta$, always larger than the value given by $(3.11)$
for ${\bf p} \neq 0$.  The spectrum for the two-body $N + \theta$
channel is therefore
$$ w_+(k) = \biggl( {p^2 \over 2M_N} - {m_\theta^2 \over 2}
\bigl( {1 \over M_\theta} + {1 \over M_N} \bigr) + {m_\theta \sqrt
{-p^2} \over M_N} \ \ , \ \infty \biggr).  \eqno(3.12)$$
\par  Let us now return to $(3.3)$ for $k^0 < 0$.  In this case,
$$ \eta_-(k) = -p^0\sqrt{{\bf k}^2 + m_\theta^2} - {\bf p}\cdot
{\bf k}   \eqno(3.13)$$
goes to $-\infty$ for $\vert {\bf k}\vert \rightarrow \infty$,
and the spectrum is unbounded from below.  The upper bound
for fixed $m_\theta$ occurs at a value of $\vert {\bf k}\vert$
for which (the second derivative is negative)
$$ \eta'_- (k) = {-p^0 \vert {\bf k} \vert \over
\sqrt{{\bf k}^2 + m_\theta^2} } - \vert {\bf p} \vert \cos\,\phi,
\eqno(3.14) $$
yielding, for $\cos \,\phi <0$,
$$ \vert{\bf k} \vert = {\vert {\bf p}\vert
 \vert\cos\,\phi \vert \over
\sqrt{{p^0}^2 - {\bf p}^2 \cos^2\,\phi}} m_\theta ,
\eqno(3.15)$$
so that
$$\bigl(\eta_-(k) \bigr)_{max} =
 - m_\theta \sqrt{{p^0}^2 - {\bf p}^2 \cos^2\,\phi} , \eqno(3.16)$$
which is further maximized by taking $\phi = \pi$:
$$ \bigl( \eta_-(k) \bigr)_{max} = -m_\theta \sqrt {-p^2} ,
\eqno(3.17)$$
  The spectrum
for the two body $N + \theta$ channel in the case $k^0 <0$
is then
$$ w_-(k) = \biggl( -\infty,\  { p^2 \over 2M_N}
- {m_\theta^2 \over 2} \bigl( {1 \over M_\theta} + {1 \over M_N}
\bigr) - {m_\theta \sqrt {-p^2} \over M_N} \biggr).
\eqno(3.18) $$
The upper bound of $w_-(k)$ is less than the lower bound
of $w_+(k)$, i.e.,
$$ (w_+(k))_{min} - (w_-(k))_{max} = 2m_\theta {\sqrt{-p^2}
\over M_N} , \eqno(3.19) $$
and there is a gap in the spectrum for every $m_\theta \neq 0$.
For the case in which $m_\theta$ can reach zero, the continuous
spectrum, including {\it both } $k^0 > 0$ and $k^0 < 0$ parts,
covers the real line $(-\infty, \infty)$.
\par Let us consider the question, for $k^0 > 0$, of whether
the ``discrete'' spectral point $p^2 / {2M_V}$ is embedded in
the continuum.  The condition that this point lie above the
lower bound of $w_+(k)$ is that
$$ {m_\theta^2 \over 2} \bigl( {1 \over M_N} + {1 \over M_\theta}
\bigr) - m_\theta {\sqrt{-p^2} \over M_N} > {p^2 \over 2}
\bigl( {1 \over M_N} - {1 \over M_V} \bigr) . \eqno(3.20)$$
Although conservation of energy and momentum imply that for
a decay process, $m_V > m_N + m_\theta $, the dimensional scale
factors $M_V,\ M_N,\ M_\theta$ may satisfy this inequality
as well; when the kinematical masses
 $\sqrt{-p^2},\ \sqrt{-(p-k)^2} $ and $\sqrt{-k^2}$ take on these
values, the classical squared proper time intervals of the
corresponding free world line $-dx^\mu dx_\mu$ become equal
to $d\tau^2$; they may also be taken as the Galilean limit of the
kinematical masses$^{12}$.  In this case,
 ${1 \over M_N}-{1 \over M_V} $ is positive.  The right hand side
  of $(3.20)$ is therefore, in this case, negative.  The
minimum value of the left hand side, at
$$ (m_\theta)_{min} = {M_\theta \over {M_N + M_\theta}}\sqrt{-p^2}
\eqno(3.21)$$
is
$$  -{p^2 \over 2} {M_\theta \over {M_N(M_N + M_\theta)}}
>0, \eqno(3.22)$$
and hence, in this case the discrete spectral point is embedded.
It is not, however, {\it necessary} to make
 the choice $M_V > M_N$; a choice of parameters can be made
  for which the discrete state lies below the minimum of
$w_+(k)$.  This would correspond to the isolated ``bound state''
case of the Lee model, for which some studies in mass
renormalization have been made$^{39}$.
\par As for the two body problem discussed in Eq. $(1.9)$ and the
following paragraph, the kinetic terms of the two body final
state can be expressed in terms of the total momentum and relative
momentum
$$ P^\mu = (p^\mu -k^\mu) + k^\mu = p^\mu \eqno(3.23)$$
and
$$ {p^\mu}_{rel} = { M_\theta (p^\mu - k^\mu) - M_Nk^\mu \over
M} = {M_\theta \over M} p^\mu -k^\mu, \eqno(3.24)$$
as
$$ w(k) = {P^2 \over 2M} + {p_{rel}^2 \over 2m}, \eqno(3.25)$$
where $M= M_N + M_\theta , \ \ m= M_N M_\theta / (M_N + M_\theta)$.
\par For the scattering system corresponding to $N+\theta
\rightarrow N+\theta$, the Lippmann-Schwinger equations provide
an effective potential for transitions in $p_{rel}$, and the
 problem may be treated as discussed in Section 1.  This will
  be done in detail elsewhere.
\par For the positive energy contributions, the process
$V \rightarrow N+\theta$ is concerned only with particles
in the usual sense, and may be represented as a process in space-
time as in fig.1.

\centerline {[fig. 1]}

\par For $k^0 <0$, however, the space-time picture appears as in
fig. 2, corresponding to Stueckelberg's representation of
pair annihilation as part of the process
 $V \rightarrow N + \theta$.

\centerline {[fig. 2]}

\noindent The corresponding physical process,
 seen in the laboratory, contains the $CPT$ conjugate of the
$\theta$ external leg, and appears as in fig. 3, corresponding to
a real $V+{\bar \theta}$ annihilation to yield the (virtual) $N$.

\centerline {[fig. 3]}

\noindent The inverse process, $ N \rightarrow V + {\bar \theta}$
corresponds to the physical decay of the $N$ particle.
For the case $n \rightarrow p + \pi^-$ discussed
 by Henley and Thirring$^{27}$, this inverse process corresponds to
$ p \rightarrow n + \pi^+$.  The difficulty that they cite,
 that in the on-shell
relativistic quantum field theories both of these processes
occur simultaneously in the interaction term of an analog
relativistic Lee model, does not occur in the model presented here,
since the two processes, although described in the context of
the same theory, occur in two different parts of the spectrum.
\par For the case $k^0 < 0$, the condition that the discrete
spectral point is embedded in the continuum is
$$ {m_\theta^2 \over 2} \bigl( {1 \over M_\theta} + {1 \over M_N}
\bigr) + m_\theta {\sqrt{-p^2} \over M_N} < {p^2 \over 2}
\bigl( {1 \over M_N} - {1 \over M_V} \bigr).\eqno(3.23) $$
If $M_V > M_N$, the right hand side is negative, and hence the
discrete point is an isolated bound state.  However, in the
physical process $N \rightarrow V + {\bar \theta}$, the scale
parameters may be chosen (as in the argument given for the
process $V \rightarrow N + \theta$) to satisfy $M_N > M_V$, in
which case the right hand side of $(3.23)$ is positive.  The
left hand side can extend to zero, and hence for this choice,
the discrete state is embedded in the continuum as well.
Since, as we have seen, the inclusion of both positive and negative
branches of $k^0$ results in a spectrum that covers the whole real
line (when $m_\theta$ reaches zero), the discrete spectral
point would then be embedded for any $M_V, \ M_N$.
\bigskip
{\bf 4. The Friedrichs Form and Conclusions}
\smallskip
\par Since the relativistic Lee model we have considered above
has sectors, it can be isomorphically mapped to a spectral model.
We write the generator of the motion in the direct integral
space $ \int_\oplus {\cal H}_p$ over the four-momenta
of the $V$ particle as (for each $p^\mu$)
$$ K_p = {\cal M}_V(p) P_0 + {\bar K}_p + K_I,\eqno(4.1)$$
where $K_I$ has matrix elements only between the continuum and the
discrete state $\phi_p$ (here, $P_0 = \phi_p \phi_p^\dagger$)
with eigenvalue ${\cal M}_V(p)$  corresponding to $ {p^2/ {2M_V}}$,
 and ${\bar K}_p $ has absolute
continuous spectrum (with, perhaps, finite or semi-infinite gaps).
This form corresponds to the canonical decomposition
$$ K_p = P_0 K_p P_0 + {\bar P}K_p {\bar P} + P_0K_p{\bar P}
+ {\bar P} K_p P_0, \eqno(4.2)$$
where ${\bar P}$ is the projection into the continuous spectrum
(at each point $p$).
\par The vertex function $f(k)$ of the
 field theoretical model can be
covariantly restricted to support in the upper light cone,
$k^0 >0$, and the spectrum (with bounded $m_\theta$) will be
therefore similar to that of the usual non-relativistic Friedrichs
model, i.e., for which the unperturbed evolution operator
has semi-infinite absolutely continuous spectrum and an embedded
discrete state with an eigenvalue corresponding to the mass of the
initial state of the decay problem (note a similar phenomenon in
structure of the Feynman propagators studied in ref. 40).
\par As we have seen, if $k^0$ can be positive or negative,
the continuous spectrum can reach $\pm\infty$; if $m_\theta$
can reach the value zero, the continuous
spectrum will have no gaps, and the spectral model is very similar
to that of Pietenpol$^{41}$; if $m_\theta$ is bounded away
from zero, there will be a
gap.  Furthermore, the discrete state may occur embedded in either
branch of the continuum or may occur within the gap.  There are
therefore a large variety of covariant models which may be
extracted from the framework given here.
\par Denoting, for each $p$, the second term on the right of
$(4.1)$ by
$$ {\bar K}_p = \int \  w(k) \vert k \rangle \langle k \vert
d^4k ,  \eqno(4.3)$$
 a calculation of the type given above, from $(1.28)$ to
$(1.35)$, results directly in the relation $(2.23)$ for the
reduced resolvent at each $p$.
\par Somwhat more abstractly, introducing a
 continuous label $\lambda$,
we may denote the continuum
generalized eigenstates by $\{\vert \lambda \rangle\}$, so
that $(4.3)$ takes the form
$$ {\bar K}_p = \int \ w(\lambda) \vert \lambda \rangle
\langle \lambda \vert d\lambda.$$
  The reduced resolvent then takes the generalized Friedrichs form
$$ h(z,p) = z - {\cal M}_V(p) -\int d\lambda \ {\vert \langle
\lambda \vert K_I \vert \phi_p )\vert^2 \over z -w(\lambda)}.
\eqno(4.4)$$
  The complex poles which dominate
the decay law (in ``proper time''), as discussed at the end
of Section 2,
can then be investigated in a way similar to the discussion
of the nonrelativistic form in Section 1.
The construction of the generalized states$^4$ of the
relativistic Lee-Friedrichs model can then be carried out.
\par We remark here that the generalized
 eigenstate in this construction corresponds to a shift is the
position of the unperturbed eigenvalue of $K_p$ from
${\cal M}_V(p)$ to the complex zero of $h(p,z)$ in the second
Riemann sheet.  This can be interpreted as the acquisition
of a complex part to the total energy-momentum of the system
$p^\mu$.  Let us label the pole position by (we neglect the
dependence of $M',\ \gamma$ on $p$ here)
$$ \zeta = -{M'^2 \over 2M_V} - i {\gamma \over 2}, \eqno(4.5)$$
so that, as for the nonrelativistic case, the decay law goes as
$e^{-\gamma\tau}$.  Then, in the center of momentum frame,
$$ {p^2 \over 2M_V} \rightarrow -{E^2 \over 2M_V} \sim \zeta, $$
or, for $\gamma \ll M'$,
$$ E \sim M' + i {M_V \over 2M'} \gamma . \eqno(4.6)$$
For $M' = M_V + \delta M_V $ ( we assume ${p^2 \over 2M_V}
\sim -{M_V \over 2} $ for the unperturbed system),
 $$ E \sim M_V + \delta M_V + i {\gamma \over 2}, \eqno(4.7)$$
where we have neglected a term of order $\gamma ({\delta M_V /
M_V})$.  The result $(4.7)$ illustrates how, in the rest frame
of the initial particle, the mass change and decay width
generated by the interaction in this covariant model
can emerge as a complex energy.
\bigskip
{\it Acknowledgements\/} I would like to thank
 I. Antoniou and S. Tasaki
for discussions, and I. Prigogine for the opportunity to
present this work at a seminar at the International
 Solvay Institutes for Physics and Chemistry at the
Free University of Brussels. I am also grateful to
  N. Shnerb, E. Eisenberg, at Bar Ilan University,
  L. Burakovsky and M. Land
at Tel Aviv University and
M. Berkooz, at Rutgers University for helpful discussions,
 and to W.C. Schieve for his hospitality at
the University of Texas at Austin, and
for very
helpful discussions at an early stage of this work.
\vfill
\eject
\noindent{\bf References.}
\frenchspacing
\bigskip
\item{1.} T.D. Lee, Phys. Rev {\bf 95}, 1329 (1954).
\item{2.} K.O. Friedrichs, Comm. Pure and Applied Math. {\bf 1},
361 (1950).
\item{3.} L.P. Horwitz and J.-P. Marchand, Rocky Mountain
Journal of Mathematics {\bf 1}, 225 (1971).
\item{4.} I. Sigal and L.P. Horwitz, Helv. Phys. Acta {\bf 51}
685 (1978); W. Baumgartel, Math. Nachr. {\bf 75}, 133 (1978);
G. Parravicini, V. Gorini, and E.C.G. Sudarshan, Jour. Math.
Phys. {\bf 21}, 2208 (1980);
T. Bailey and W.C. Schieve, Nuovo Cimento {\bf 47A}, 231 (1978);
Physics {\bf 78}, Berlin (1978); A. Bohm, {\it Quantum Mechanics:
  Foundations and Applications}, Springer, Berlin (1986);
 A Bohm, M. Gadella and
G.B. Mainland, Am. Jour. Phys. {\bf 57}, 1103 (1989).
\item{5.} A. Bohm, {\it The Rigged Hilbert Space and Quantum
Mechanics}, Springer Lecture Notes on Physics {\bf 78}, Berlin
(1978); I.M. Gel'fand and N. Ya. Vilenkin, {\it Generalized
Functions}, vol 4, Academic Press, N.Y. (1964).
\item{6.} I. Antoniou and S. Tasaki, J. Phys. A: Math. Gen
{\bf 26}, 73 (1993); H.H. Hasegawa and W.C. Saphir, Phys.
Lett. {\bf 161A}, 471 (1992); Phys. Rev. A {\bf 46}, 7401 (1992).
\item{7.} I. Antoniou and I. Prigogine, Physica A {\bf 192},
443 (1993), and references therein.
\item{8.} L.P. Horwitz and C. Piron, Hev. Phys. Acta {\bf 46},
316 (1973); R. Fanchi, Phys. Rev D {\bf 20}, 3108 (1979);
C. Dewdney, P.R. Holland, A. Kyprianides
and J.P. Vigier, Phys. Lett. {\bf A 113}, 359 (1986); Phys.
Lett. {\bf A 114}, 444 (1986); A. Kyprianides, Phys. Rep.
{\bf 155}, 1 (1986).
3108 (1979).
\item{9.} E.C.G. Stueckelberg, Helv. Phys. Acta {\bf 14},
 322, 588 (1941); J. Schwinger, Phys. Rev. {\bf 82}, 664 (1951);
R.P. Feynman, Rev. Mod. Phys. {\bf 20} 367 (1948); Phys. Rev.
{\bf 80}, 440 (1950).
\item{10.} L.P. Horwitz, Found. of Phys. {\bf 22}, 421 (1992).
\item{11.} See R. Arshansky and L.P. Horwitz,
 Found. of Phys. {\bf 15},
701 (1985) for a discussion of localization in space and time,
and the associated Newton-Wigner and Landau-Peierls operators.
M. Usher and L.P. Horwitz, Found. Phys. Lett. {\bf 4}, 289 (1991),
have discussed the causal properties from the point of view taken
by G.C. Hegerfeldt, Phys. Rev. D {\bf 10}, 3320 (1974); Phys.
Rev. Lett. {\bf 54}, 2395 (1985); Nucl. Phys.
 B {\bf 6}, 231 (1989).
\item{12.} L.P. Horwitz, W.C. Schieve and C. Piron, Ann. Phys.
{\bf 137}, 306 (1981).
\item{13.} L.P. Horwitz, S. Shashoua and W.C. Schieve,
Physica A {\bf 161}, 300 (1989).
\item{14.} L. Burakovsky and L.P. Horwitz, Physica A {\bf 201},
666 (1993); ``Galilean limit of equilibrium relativistic
mass distributions,'' TAUP 2081-93, to be published;`` Equilibrium
relativistic mass distributions for indistinguishable events,''
TAUP 2115-93; L. Burakovsky and
L.P. Horwitz,``Independence of specific heat on mass distribution
for large c,'' TAUP 2141-94.
\item{15.}  L. Burakovsky, L.P. Horwitz and W.C. Schieve,
``Statistical mechanics of relativistic degenerate Fermi gas I.
Cold adiabatic equation of state,'' TAUP 2136-94.
\item{16.} L. Van Hove, Physica {\bf 21}, 901 (1955); {\bf 22},
343 (1956); {\bf 23}, 441 (1957).
\item{17.} R.I. Arshansky and L.P. Horwitz, Jour. Math. Phys.
{\bf 30}, 66, 380 and 213 (1989).See also R. Arshansky and
L.P. Horwitz, Phys. Rev. D {\bf 29}, 2860 (1984).
\item{18.}M.C. Land, R.I. Arshansky and L.P. Horwitz, to be
published, Found.of Phys.
\item{19.} M.C. Land and L.P. Horwitz, ``The Zeeman effect for
the relativistic bound state,'' TAUP-2150-94, to be published.
\item{20.} L.P. Horwitz, Jour. Math. Phys. {\bf 34}, 645 (1993).
\item{21.} R. Faibish and L.P. Horwitz, ``Non-compact dynamical
groups for the solution of bound state problems in relativistic
quantum theory I: The relativistic harmonic oscillator in
$1+1$ dimensions,'' TAUP 2025-93, to be published in Jour. Group
Theory and Applications; ``Dyanmical groups of the relativistic
Kepler problem and the harmonic oscillator,'' vol. II, p. 450,
Anales de Fisica, Monografias, vols. I, II, ed. M.A. del Olmo,
M. Santander and J.M. Guilarte, CIEMAT-RSEF, Madrid (1993),
\item{22.} E.P. Wigner and V.F. Weisskopf, Zeits. f. Phys.,
{\bf 63}, 54 (1930); {\bf 65}, 18 (1930).
\item{23.} P. Exner, {\it Open Quantum Systems and Feynman
Integrals},  D. Reidel, Boston (1985); A. Bohm, {\it Quantum
Mechanics: Foundations and Applications}, Springer, Berlin
(1986). See also L.A. Khalfin, Dokl. Akad. Nauk. USSR,
{\bf 115}, 277 (1957), for early studies of the long and
short time behavior.
\item{24.} L.P. Horwitz and C. Piron, Helv. Phys. Acta {\bf 66},
693 (1993).
\item{25.} See, for example, J.-P. Marchand and L.P. Horwitz,
Rocky Mountain Jour. Math. {\bf 1}, 225 (1971)
\item{26.} N. Bleistein, R. Handelsman, L.P. Horwitz and
H. Neumann, Nuovo Cimento {\bf 41A}, 389 (1977).
\item{27.} E. Henley and W. Thirring, {\it Elementary Quantum Field
Theory}, McGraw Hill, New York (1963).
\item{28.} B. Zumino, in {\it Lectures on Field Theory and
the Many Body Problem}, ed. E.R. Caianiello, p. 37, Academic
Press, New York (1961).
\item{29.} L. Van Hove, second of refs. 16.
\item{30.} D. Saad, L.P. Horwitz and R.I. Arshansky, Found. Phys.
{\bf 19}, 1126 (1989); N. Shnerb and L.P. Horwitz, Phys. Rev A
{\bf 48}, 4068 (1993).
\item{31.} J. Frastai, in preparation.
\item{32.} J. Schwinger, Phys. Rev. {\bf 82}, 664 (1951).
\item{33.} W. Pauli and F. Villars, Rev. Mod. Phys. {\bf 21},
434 (1949).
\item{34.} See, for example, J.M. Jauch and F. Rohrlich,
{\it The Theory of Photons and Electrons}, 2nd ed.,
 Springer, New York
(1976); C. Itzykson and J.-B. Zuber, {\it Quantum Field Theory},
McGraw Hill, New York (1980).
\item{35.} See, for example, L. de Broglie,
 {\it Les incertitudes d'Heisenberg
et l'interpr\'etation probabiliste de la m\'ecanique
ondulatoire}, preface by Georges Lochak, Gauthier-Villars,
Paris (1982).
\item{36.} K. Haller and R.B. Sohn, Phys. Rev A {\bf 20},
1541 (1979); K. Haller, Acta Phys. Austriaca {\bf 42}, 163
 (1975); K. Haller, Phys. Rev D {\bf 36}, 1830(1987).
\item{37.} L.P. Horwitz, C. Piron and F. Reuse, Helv. Phys. Acta
{\bf 48}, 546 (1975); C. Piron and F. Reuse, Helv. Phys. Acta
{\bf 51}, 146 (1978);L.P. Horwitz and R. Arshansky, J. Phys. A:
Math. Gen. {\bf 15}, L659 (1982); A. Arensburg and L.P. Horwitz,
Found. Phys. {\bf 22}, 1025 (1992).
\item{38.} P.A.M. Dirac, {\it The Principles of
 Quantum Mechanics}, 3rd ed., Oxford University
  Press, London (1947).
\item{39.} G. K\"all\'en and W. Pauli, Mat. Fys. Medd.
Dan. Vid. Selsk. {\bf 30}, 1 (1955).
\item{40.} L.P. Horwitz and Y. Lavie, Phys. Rev D {\bf 26},
819 (1982).
\item{41.} J.L. Pietenpol, Phys. Rev. {\bf 162}, 1301 (1967);
I. Antoniou, J. Levitan and L.P. Horwitz, J. Phys. A: Math.Gen.
{\bf 26}, 6033 (1993).

\vfill
\eject
\centerline {\bf Figure Captions}
\bigskip
\item{fig. 1.} The process $N \rightarrow V + \theta(k^0 >0)$
\item{fig. 2.} The process $N \rightarrow V + \theta(k^0 <0)$
\item{fig. 3.} The process $N + {\bar \theta}(k^0 >0) \rightarrow V
{\rm (virtual)}$
\vfill
\eject
\end
\bye